# A View on Edge caching Applications

D. Antonogiorgakis, A. Britzolakis, P. Chatziadam, A. Dimitriadis, S. Gikas, E. Michalodimitrakis, M. Oikonomakis, N. Siganos, E. Tzagkarakis, Y. Nikoloudakis, S. Panagiotakis, E. Pallis, E.K.Markakis
*Department of Electrical and Computer Engineering*
*Hellenic Mediterranean University*
*Heraklion, Crete – Greece*

Emails: {mtp182, mtp186, mtp198, mtp199, mtp202, mtp185, mtp183, mtp194, mtp196}@edu.hmu.gr, {nikoloudakis, spanag, epallis, emarkakis}@staff.hmu.gr

***Abstract*—** Devices with the ability to connect to the internet are growing in numbers day by day thus creating the need for a new way of managing the way the produced traffic travels through data networks. Smart Cities, Vehicular Content Networks, Healthcare and Virtual Reality Videos are a few examples that require high volume data while maintaining low latency. Edge caching practices are a promising solution in such cases in order meet the requirements of low latency in high volume traffic. This paper is a survey on four indicative areas, Smart Cities, Vehicular Content Networks, Healthcare and Virtual Reality Videos that make use of edge caching.

## 1. Introduction

The emerging growth in numbers of small devices with the ability to connect to the internet, has created the need for a new way of managing the way the produced traffic travels through data networks. This new state of things, has established a need for uninterrupted, lossless and without any delay, communications. There are three major ways with whom these issues can be addressed. Edge Caching, Edge Computing and in some cases Software Defined Networking. We will focus on edge caching, by reviewing recent papers that offer solutions for improving edge caching in various applications.

One major category of applications, which can benefit from edge technologies, is the category of medical applications. Medical applications have a direct impact on humans because the data they manage comes from either their physical condition, their medical condition or their pharmaceutical care. Of course, this type of applications is also influenced by many factors that may pose serious problems to individuals who rely on a health care system. The increase of mobile and wireless devices has the effect of overloading the network, due to the fact that we require fast communication and transmission of a large volume of data (4K images), or in many cases we need real-time supervision and monitoring, especially for emergency situations.

Virtual reality applications are emerging due to the fact of offering 360 viewing experience. These applications are very demanding in aspects of capacity and computational capabilities as well as delay sensitive.

Although the area of applications is vast, in this paper we focus only on four indicative areas, Smart Cities, Vehicular Content Networks, Healthcare and Virtual Reality Videos. The rest of this paper is as follows: Section 2 presents applications of edge caching, in IoT environments and in Section 3, we focus on applications in Vehicular Content Networks and explore the concepts of energy in the matter. In Section 4, we present edge caching applications in healthcare and in Section 5, we present edge caching applications in Virtual Reality videos. Finally, section 6 includes concluding remarks and Section 7 the references.

A summary of existing survey and research publications on edge Caching Applications is shown in Table 1, while a legend of the publication categories is shown in Table 2.

*Table 1 – Summary of existing survey and research publications on Edge Caching Applications.*

| Aspects | Categories | Publication | Contributions |
|---|---|---|---|
| Smart City - Internet of Things | A, F | [1] | An introduction of an efficient caching policy by deploying a deep reinforcement learning (DLR) algorithm |
| | A, C, G | [2] | A solution for the dissemination of data produced by IoT devices by using IOT devices with the ability to cache data and exchange data with others. |
| | A, D | [3],[39] | Presentation of a value-based caching framework for Information Centric Sensor Networks that takes into account the main functional parameters and uses a new component called Local Cognitive Node. |
| | A, D | [4] | An effort to overcome the limitations of current networks by introducing a "Smart Collaborative Caching" scheme (SCC) by applying high-level ICN principles within the fog computing paradigm. |
| | A, D | [5] | A proposal of a caching strategy along with a freshness mechanism under Named Data IoT Networks |
| | A, B | [6] | A solution for the joint problem of optimization of caching, computing, and radio resources allocation in the fog-enabled IoT. Deep learning is used in different stages, in order to minimize the end–to-end service latency |
| | A ,B, C | [7] | Introduction of a novel AI-enabled smart edge with heterogeneous IoT architecture, a new architecture and an algorithm for wireless communication, collaborative cache filtering and computation offloading models. |
| | A, D | [8] | An integrated framework that can enable dynamic orchestration of networking, caching, and computing resources to improve the performance of applications for smart cities. |
| Vehicular Content Networks | A,C | [9] | Analyzes the features of a vehicular content network, addresses the issue of the limited storage at RSUs and proposes a model that determines whether and where to obtain the replica of content when the moving vehicle requests it |
| | A,C, I | [10] | Propose an air-ground integrated vehicular network architecture, whereby the High Altitude Platforms broadcast popular contents to vehicles a a-priori to requests while the ground roadside units (RSUs) provide services on demand through unicast. |
| | A, I | [11] | A proposal for a prefetching policy for edge caching, which is based on specifying the End Nodes that a vehicle will meet on its route and estimating the time that will be in the range of each End Nodes |
| | A, C, D, H | [12] | An approach that emphasizes in the orchestration of VCN in a higher level with structure and management based on SDN and NFV with ICN techniques and deep reinforcement learning. |
| | A, F | [13] | Another aspect in edge caching that is based on in-vehicle caching with a dynamic distributed storage relay mechanism. |
| | A, C, F | [14] | An alternative method with reverse CDN for Content Caching vehicle applications. |
| | A, H | [15] | An SDN implementation, to reduce the delay for the data traffic in transportation systems. |
| | D, F, I | [16] | A proposal for resources in ITS for the future, using static nodes. |
| | A, G, K | [17] | An analysis of energy efficiency in 5G systems. |
| | A, G, K | [18] | A cache-enhanced Mobile Edge Computing - MEC system is presented which focuses on task- caching on the edge servers in order to avoid duplicate data transmissions. |
| | A, G, K | [19] | An analysis on content caching techniques is performed in order to reduce congestion as well as latency. |
| Healthcare | A, E | [20] | Developing E-health system for emergencies based on fog-edge and CNN architectures |
| | A, C | [21] | An analysis of collaborative cache allocation and computation offloading in mobile edge computing, for edge servers creation to manage medical images |
| | A, C | [22] | An analysis of medical monitoring systems for people with diabetes. |
| | A, C, G | [23] | A study of edge caching in 5G networks and proposing an architecture using small cell cloud and microcell cloud based on SDN based mobile network. |
| | A, F | [24] | A study and development of a system for transmitting visual and non-visual medical information through edge caching and Vision Sensing (DVS) technologies |
| | A, C | [25] | An analysis of the properties of a Fog e-health gateway and report the key pillars of it |
| | A, G | [26] | An analysis of mobile caching-enabled small-cells for delay-tolerant e-Health apps |
| | A, C | [27] | Developing a fog computing system that enables the storage and communication of various medical data such as phonocardiogram, electrocardiogram and speech data |
| | A, J | [28] | A comparison of medical applications and architectures so that each time they find the optimal microprocessor architecture for each application |
| Virtual Reality Videos | A, B, G | [29] | A comprehensive review on the state-of-the-art research efforts on mobile edge networks that includes definitions, architectures, communication techniques and advantages. |
| | A, B, C | [30] | An analysis of a specialized cache system, named Cachier, for image-recognition applications that yields enhanced speed and responsiveness. |
| | A, C | [31] | A description of a communications-constrained MEC framework that promises to reduce communication-resource consumption by fully exploiting the computation and caching resources at the mobile VR device. |
| | A, C, G | [32] | A presentation of a novel MEC-based mobile VR delivery framework by jointly utilizing the caching and computing capacities of the mobile VR device thus achieving low latency and low bandwidth utilization. |
| | A, F | [33] | An exploitation of two novel caching strategies that mine user/group interests to improve caching performance at net- |
| | | | work edge thus improving performance for high-throughput low-delay applications, such as virtual reality, augmented reality and Internet-of-Things. |
| | A, G | [34] | An analysis of an optimization framework for delivery of VR/AR 360°-navigable videos to 5G VR/AR wireless clients in future cooperative multi-cellular systems thus maximizing the aggregate reward for caching/computing resources. |
| | A, D, G | [35] | An analysis of the impact of AR/VR on the networking Infrastructure with a special focus on 5G Information-Centric wireless networks that promise to significant benefits to better support AR/VR. |
| | A, G | [36] | An exploitation and case study presentation that underline the importance of VR technology as a disruptive use case of 5G (and beyond) harnessing the latest development of storage/memory, fog/edge computing, computer vision and artificial intelligence. |

*Table 2 - Publication Categories Legend.*

| Categories | Identifier |
|---|---|
| Content Caching | A |
| Task Caching (offloading) | B |
| Architectural Approaches | C |
| ICN | D |
| CCN | E |
| Data Driven | F |
| 5G Mobile Networks | G |
| Software Defined Networking | H |
| Prefetching Content | I |
| Benchmark | J |
| Energy Efficiency | K |

## 2. Edge caching in Internet of Things

The authors in paper [1] try to invent an efficient caching policy by deploying a deep reinforcement learning (DLR) algorithm. The purpose of this algorithm is to refresh cached data by newer ones when necessary. In doing so, they decrease the amount of network utilization from IoT devices to the edge caching node(s). This method requires the addition of two fields in each data item. The timestamp field, indicating the time generation of data and the lifetime field, indicating the duration for the validity of the data item. The authors identify three case scenarios for an edge caching node upon receiving a request from an application. The first one is when the node has the requested item and its state is “fresh”, therefore no new data arrival is needed. The second is when the node has the item, but the lifetime indicates that it is not valid anymore, consequently a new “fresh” data item has to replace the existing. Finally, if there is no associated data item with the request, the node has to fetch a new data item. The question that rises here is which cached data item to replace. In order to decide, the writers introduced two functions representing communication cost and freshness loss cost. Using those, they formulated the cache replacement problem as a Markov decision process. They adopted a deep reinforcement learning algorithm to discover the optimum solution, without any knowledge of the prior state of the system. They then performed a simulation of the system using their algorithm of neural networks in comparison with two baseline edge caching policies. LRU: The new data item replaces the one least requested. LFF: The new data item replaces the one which is least “fresh”. The results showed that the policy proposed by the writers outperformed LRU and LFF for various parameters regarding the networking setup.

Antonino Orsino et al. in paper [2] try to find a solution for the dissemination of data produced by industrial IoT devices. The volume of data to be transmitted in an industrial environment most of the time is small. However, when it comes to monitoring devices that produce video data, this is no longer the case. Fifth Generation (5G) technology is a promising solution for fast seamless transmission. However, taking into account that 5G operates in millimeter-wave frequencies leads us to the fact that trouble-free communications may be challenging. The propagation of wireless signals is often obstructed by walls, machines, humans even by other IoT devices coexisting in the same area. It becomes obvious that it is imperative to find a solution addressing these problems. Contemporary IoT devices, are capable of caching contents locally. In addition, some IoT devices have the ability to exchange data with others such as segments of video files, performing a device to device communication (D2D). The authors suggested three modes for a device transmitting data “Direct Push”, “Store and Push” and “Forward and Push”. In “Direct Push”, a machine that finds itself in good radio link conditions decides to immediately transmit data instead of caching them. Then, the edge node caches the data. The second mode “Store and Push” applies if radio link conditions are not good enough for a trustworthy transmission. In this case, the device decides to cache data locally until wireless network quality is sufficient. The final approach “Forward and Push”, employs the help of a proximal device that will receive data and forward them to the edge cache node. The first mode is the fastest with zero delay if the wireless network conditions are excellent at all times. This is not the case, so it is up to the second mode to serve alternately when needed, the process. The second mode however, can introduce major delays to the data propagation, whenever the device experiences bad radio link connections. To overcome this limitation, the writers introduced a "smart data dissemination strategy" which they named "predictive dissemination". The network conditions are known due to the creation of a model map of a specific area of the factory. This LoS map has information regarding the network states in every point of interest. The devices are making decisions (between the three transmitting modes) taking into account the LoS map information. The writers proved with the help of a simulation model that their proposal offers the optimal solution in comparison with the exclusive use of mode "Direct Push" and combined mode "Direct Push" and "Store and Push". Particularly their proposal outperformed the other two modes in the fields of “acquisition delay”, “undelivered contents due to link data rate” and “undelivered content due to blockage”.

Information Centric Networks is a new advent in modern network architecture that overcomes the traditional scheme of location-based data transfer and focuses on the content itself. This architecture will maximize the ability to utilize in-network caching as an important factor of the network solution. Wireless Sensor Networks (WSN) is a key component of IoT and a source of a huge amount of information. In his paper, Fadi Al-Turjman [3] propose a new framework of “cognitive caching”, for in-network caching, designed for the future WSN. In his framework called CCFF, he describes a model for refreshing the cached data of fog nodes, by using a replacement policy that analyzes data in order to specify which of them need to be removed from the cache. Reasoning and learning methods are used to identify the regularly requested and most substantial data. In this model, a new component –a gateway- is proposed as part of the edge system, called Local Cognitive Node (LCN) with the role of maintaining information related to data popularity and network status. A typical WSN infrastructure consists of relay nodes (RN) and sink nodes (SN) that forward data to the cloud services. The ICSN data traffic model, fit in three categories: periodic, on demand and on an event (emergency) created data and have various Quality of Information values, related to the data consuming service. Those complex gateways are exchanging information with WSN components such as RN and SN and provide data replies to the users. Having awareness of both the network status and the requirements of the services, along with the embedded cognitive logic, the gateway can handle user demands with greater performance. Additional cognitive logic is based on a learning component that uses a fast heuristic search algorithm and reasoning component that uses a variance of the Analytic Hierarchy Process model, to find the best available data paths. The algorithm takes into account delays, age, popularity, and trust factors, with adjustable weights to optimize performance and it is executed by every node that its cache memory is getting full. Thus cached data refreshment is a complex decision that uses multiple criteria. Finally, the proposed framework is compared in a simulation environment using NS3 and various datasets, with other content, location and functionality-based caching methods that showed better efficiency and scalability.

The writers in [4] deal with Information Centric Networking and its application in IoT. According to them with the help of Information - Centric Networking, we can overcome the limitations of current networks and catch up with the latest demands. For doing so they proposed a "Smart Collaborative Caching" scheme (SCC). They achieved that by applying high-level ICN principles within the fog computing paradigm. Actions, including caching, and the relevant algorithms for this purpose have been introduced and presented. They set up a validation environment to compare performance differences between four solutions (ICN SCC-E, ICN SCC, ICN flooding and IP network). Their main goal was to decrease the number of packets within the network as well as the transmission latency. Results showed that ICN flooding and IP network performed rather poorly in comparison to ICN SCC-E and ICN SCC.

The “freshness” of data produced by IoT devices is the main concern of the writers in paper [5]. The problem relies on how often the data stored in edge caching needs to be replaced by newer ones. The authors present a mechanism with whom they try to manage the freshness problem and call it "event-based freshness". The algorithm compares the lifetime of the data stored in the cache with the lifetime of the corresponding data produced at the IoT device. This way they can determine if the producer has newer data than the ones already stored at edge caching. Along with that, they introduce a new caching strategy they named "consumer-cache". The main idea is to keep as low as possible the number of caching nodes and to have some privileged gateways directly connected to consumers. They also placed caching nodes "on-path" meaning that they are positioned in the network topology rather than in some central fixed place. Creating a simulation scenario the writers compared their proposed caching strategy (consumer-cache) with LCE (Leave Copy Everywhere), LCD (Leave Copy Down), Prob (a randomized version of LCE), Btw and edge-caching. The results showed that the consumer-cache strategy outperforms all the above schemes. In more detail, it showed less server hit ratio, the minimum number of hops and evictions and decreased response latency. It also had the ability to provide fresher data to consumers. Finally, they also measured a factor introduced and named by the writers, the "cache cost". This factor takes into account the trade-off between caching overhead and data availability. Consumer-cache performed better regarding this scope too.

Y. Wei et al. [6] proposed a combined approach that optimizes not only caching, but offloading computational tasks and radio resources allocation. To be able to provide networks, with the desired low latency performance and meet the bandwidth requirements of fog enabled IoT, a synergy of the above techniques is discussed in detail in a joint framework. In the network architecture, fog nodes are connected to edge routers and provide wireless connectivity to the user equipment. Edge routers are aggregating data traffic and provide high-speed connectivity to content sources, online service and cloud storage. Caching can be very efficient when implemented in edge routers with internal storage and embedded functionality, to decide which content to store. Optimizing decision making strategy is critical for efficient caching. Using the popularity of the content as the criterion for caching decisions has performance limitations since there is a lack of ability to make predictions and be more adaptive to changing content. Using Machine Learning (ML) is a common technique that leads to an improved cache hit probability. In contrast with other approaches, in their paper, they propose a Deep Reinforcement Learning (RL) with the use of actor-critic. They embed the popularity of the content, as a factor in the reward function of the RL algorithm. Edge router can act as the agent of the algorithm and be able to take its own caching decision. They utilize two deep neural networks as actors. Using the iterations of improvement-evaluation steps of the action–critic model and optimize the learning performance. In the evaluation of their model they showed, that better caching performance is achieved, and results are even better when caching is combined with computing and radio resource allocation.

In a similar context, the collaboration of communication, computation and caching on edge nodes with AI capability is discussed in the paper by Yixue Hao et al. [7]. The model called Smart Edge CoCaCo is suitable for IoT infrastructures, which are in general characterized by great diversity of protocols, devices, and applications. The model introduces a new heterogeneous architecture for IoT that combines the above techniques, in order to improve the hardware infrastructure and achieve reduced delay and improved user QoE. In Smart Edge CoCaCo algorithm, computation offloading from different users is taking into account for the caching filtering decision that utilizes a content-based scheme. As a joint method, the algorithm involves many communication parameters such as delays, data volume etc. and computational factors such as complexity but also available edge node and cloud resources. For the evaluation of the proposed algorithm, they performed experiments in an emotion recognition scenario for an increasing number of users, which showed better performance than using traditional cloud techniques.

The writers in paper [8] are dealing with software-defined networks and virtualized networks that include mobile edge computing and edge caching. Specifically, they study the application of these technologies in smart city environments. A software-defined network offers a great number of advantages, which is why the researchers use one in their implementation. They also try to benefit from the use of Information-Centric Networks that are able to minimize retransmissions. In addition the use of mobile edge computing and caching, reduce network latency, therefore, increase the quality of offered services. The innovation in this research is the introduction of reinforcement learning, which is an important part of machine learning. With the use of reinforcement learning, a deep Q-network is created that yields an optimal action for each given data. Keep in mind that input data are of very high complexity. The main issue here was the efficient allocating of resources. With the use of simulation experiments, the authors proved that the proposed method can fulfill the above requirement both effectively and efficiently.

## 3. Edge caching: Concepts in vehicle networks, transportation issues and energy efficiency

### 3.1. Edge caching in Vehicular Content Networks

With car interconnection various applications can be used that will help to increase safety, improve driving conditions, reduce accidents and traffic problems. Passengers will be able to enjoy entertainment services like movies, games, etc. All these applications require direct access to the necessary data through the network. An effective way for reducing latency is the use of edge caching techniques in Vehicular Content Networks (VCN).

As Su et al. [9] describe, a VCN consists of the vehicles and the Road-Side Units (RSUs). An On-Board communication Unit (OBU) with limited caching storage has been added on vehicles. The RSUs are deployed along the road and extend the caching capability at the edge of the network, as they host copies of the data content. When an application needs content it queries it in neighboring RSUs and OBUs and sequentially to the cloud. However edge caching in VCN has some special features in relation to other fields of edge caching. The network is more dynamic as the VCNs users are mainly fast-moving vehicles that may move from one's RSU range to another's before they acquire the queried content. Apart from that, the moving cars that distribute their cache content to other network elements may draw away from the other elements before they complete the content transmission. VCN cache the most popular content and the definition of popularity is crucial for its effectiveness, as it varies depending on the locality. Hence, a fast car which crosses multiple geographical locations will not receive the desirable content faster because that may not be popular in other regions. This car will also influence the content popularity of the area that it moves in. The popularity of content also changes dynamically according to the requests, so a popular content will stop being popular when everyone has cached it.

### 3.2. Edge Caching Architectures that extend VCN

Some approaches about how VCN could be optimized through edge caching, are the following:

#### 3.2.1. Air-Ground integrated networks

Air-ground integrated networks may enhance the capabilities of VNC by providing an umbrella that will help better communication among the RSUs and also between the RSUs and the cloud. They are based on High Altitude Platforms (HAP) which are balloons/crafts objects residing in the stratosphere. Zhang et al. [10] propose an air-ground architecture for VCNs that benefits from HAPs by reducing the latency and the traffic and enhances the access rate of RSU. The HAPs classify the content by its popularity and broadcast the most popular content to vehicles, while the RSUs unicast to vehicles on demand. This architecture poses challenges to network management because of the variation of the resources and the traffic demands. A solution that Zhang et al. [10] suggest is the oriented-service network slicing method, where a network slice is created on the top of physical resources which guarantees the QoS. Proactive content pushing reduces duplicated transmissions, improves the efficiency and QoS.

#### 3.2.2. In-Order Delivery for Connected Cars with prefetching in Edge Caches

The fact that cars are constantly moving, increases the complexity of caching policies at the Edge Nodes (EN) of a VCN. Mahmood et al. [11] propose RICH (RoadsIde CacHe) a prefetching policy for edge caching. Rich is based on specifying the ENs that a vehicle will meet on its route and estimating the time that will be in the range of each EN. They consider into fixed-size chunks strongly correlating the download process at each EN, due to the in-sequence chunk delivery. RICH is using a centralized architecture and protocol where a central module manages the content that will be stored in each EN. An analytical model is used for predicting the probability of downloading a chunk from a given EN and a Prefetcher module instructs, in advance, each EN on which part of the content to cache.

#### 3.2.3. Deep Reinforcement Learning Approach in Vehicular Networks with SDN and ICN

Another approach that emphasizes in the orchestration of VCN in a higher level is proposed by He et al. [12], who envision a framework with structure and management are based on Software Defined Networking and Network Function Virtualization. Regarding the content location, they adopt information-centric networking (ICN) techniques. Content is cached at network edge nodes, increasing the network efficiency, by reducing the latency and the duplicate content transmission. This approach uses a high complexity system which makes it hard to solve the resource allocation problem with traditional methods. For this, He et al. [12] propose the use of deep reinforcement learning for the choice of optimum result for each input.

#### 3.2.4. In-Vehicle Caching via Dynamic Distributed Storage Relay in Vehicular Networks

Hu et al. [13] present another aspect in edge caching. They propose a framework that is based on in-vehicle caching (IV-Cache). High mobility of vehicles causes data loss when a vehicle leaves a certain area. The proposed framework retains the data in a designated area by relaying the data from the vehicles that abandon the area to vehicles that arrive, at the entrance/exit of the area. This is achieved with a dynamic distributed storage relay ($D^2SR$) mechanism, which guarantees the maintenance of cached data with a high probability within the required time.

### 3.3. Edge Caching for Transport in Smart Cities Reverse Content Distribution Network

It is essential for a smart city to constantly upgrade its transport infrastructure in terms of technology. Transferring video content on vehicles is something that will concern computer engineers in the following years to come. Mustafa et. al [14] present an rCDN (reverse content distribution network). This reverse content distribution network reverses the CDN function, the network transfers content (video recordings) from vehicles and various other static devices (cameras), to the fog and then to the cloud (many in one). With this technique, the videos are dynamic (analyzed, compressed, migrated etc.) and can be made available from the fog to either cloud (rCDN) or another vehicle (CDN). The main purpose is to reduce data transfer to the cloud and the distribution of video content from the fog to vehicles, in order to improve the overall quality of the network.

#### 3.3.1. Utilization of The Spare Network Resources Using SDN

Li et al. [15] presented an implementation using SDN to reduce data traffic for vehicular networks in smart cities. Data traffic is categorized in delay-tolerant and delay-sensitive that each of them has different requirements. The delay-tolerant data is transported between the vehicles, because of the usage of the SDN without interfering with the nodes, as result of this, we have a decreasing in the data traffic of the network. Furthermore, this programmable control determines the place (cloud, vehicle and node) that the computing tasks will execute. It's important, that the control determines the place that the data will be saved, in order to help the flow of the network.

#### 3.3.2. Lookup Service for resources in an ITS system

ITS or Intelligent Transportation Systems in the future will be hard to find network resources. For this reason, Tanganelli et al. [16] present a distributed lookup service, based on Fog computing, that can find resources. The important is that this service, can adapt to the available vehicles and static nodes. Because of the network changes, as the vehicles move, the available resources are different time to time. The distributed hash table (DHT) support the range queries and multi-attribute and make the service efficient.

### 3.4. Energy efficiency in mobile edge caching

Authors in [17], present energy efficient techniques in 5G communication wireless systems. These systems can be severely overwhelmed by edge caching as well as transport networks. A combination of these technologies may result to a great amount of cost for installation as well as resource utilization. Furthermore, the evaluation of the cost impact cannot be measured with traditional performance metrics, both for spectral and energy efficiency. Authors present energy efficient techniques ($E^3$), where their main idea is to take spectral efficiency (SE) and energy efficiency (EE) costs into account, when different technologies are used in 5G systems. The authors point out that 5G systems must have the ability to handle higher amount of user generated data rate (10 to 100 times). Furthermore, ensuring reduced end-to-end latency (at least five times) has to be taken into account for devices. In order to overcome those issues, several networking architectures have been proposed as respective solutions in 5G systems such as Cloud radio access networks (C-RANs), Heterogeneous C-RANs (H-CRANs) and Fog-computing-based radio access networks (F-RANs).

When combining edge caching in transport networks (which are composed by backhaul and fronthaul X-Haul) it has to be a careful consideration, on the way of reducing the total costs of the network overhead, in an overall 5G network as well as its deployment, management and operation. Moreover, the traditional metrics of spectral (SE) and energy efficiency (EE) tend to be difficult to evaluate. Authors realize the lack of an advanced performance metric proposal in order to evaluate in total both energy efficiency (EE) and the cost induced by various advanced components of 5G systems. Yan et al conclude, that SE and EE are not a simple trade-off relation and further investigation has to be made especially in 5G systems where this trade-off relation has to be revised as well as its impact cost. Furthermore, authors provide feasible performance metrics in order to characterize flexibility, interoperability, and scalability in 5G systems as well as to show the comprehensive gains. Specifically, it proposes a performance metric called as energy efficiency ($E^3$) as a main idea to take EE and cost into account. Based on this proposed metric, impacts from combination between transport networks (X-Haul) and edge caching are characterized. The results show that there are ways to optimize energy efficiency ($E^3$) performance in 5G systems by deploying transport network (X-Haul) strategies, advanced edge caching or even a joint optimization of both transport networks (X-Haul) and edge caching strategies.

As the study of this paper suggests, edge caching and transport networks (X-Haul) are important technologies on F-RAN-based 5G systems, although they tend to conflict with themselves affecting their overall performance. Despite the fact that edge caching techniques help to ease the pressure on transport networks (X-Haul), its support for more transmissions will eventually decrease the edge caching resources. The study also shows a comparison between SE, EE and $E^3$ metrics. The numerical results reveal that the optimal $E^3$ performance cache size will be affected by the change in X-Haul. Consequently, joint consideration of both transport networks (X-Haul) and edge cache strategies deserves attention while optimizing $E^3$.

A cache-enhanced Mobile Edge Computing - MEC system is presented in [18] which focuses on task- caching on the edge servers in order to avoid duplicate data transmissions. Vu et al suggesting that mobile devices have limited amount of resources. To tackle the problem, some offload computational models were recommended such as mobile edge computing mobile, cloud computing etc. Such computing models can ease the burden of mobile devices by migrating a part or even all of the application to powerful cloud servers. Furthermore, small portable devices (such as mobile phones, tablets etc.) may have limited amount of hardware capabilities causing delay as well as energy on offloading. In addition, transmissions with massive amount of data can be a great challenge on the access network. Researchers propose edge caching as a solution due to the fact that, caching popular content at the edge servers, may result in delay reduction as well as energy consumption minimization. In particular, the term task caching is referred to the caching of application program as well as additional data (namely task data) on the edge servers. In order to optimize the energy efficiency, of the cache enhanced MEC system, they formulated a problem that can optimize both cache policy and resource allocation of computation/communication. This is important in order to reduce energy consumption on mobile devices according to its hardware requirements. They also propose an efficient iterative algorithm for solving mixed integer optimization problem.

Another interesting study is presented in paper [19] where an analysis on content caching techniques is performed in order to reduce congestion as well as latency. Specifically, Liu et. al propose an energy efficient performance project for cache assisted Content Delivery Networks - CDN with wireless backhaul. Specifically, users have options such as storing their data in their local cache when transmission signal is designed. Notably, the researchers of this study perform an investigation on energy efficiency performance in cache-assisted content delivery networks. Users request content from a data center through a wireless backhaul. During the first phase they analyze energy consumption using the two main caching strategies, coded and uncoded. In uncoded caching strategy, content is separately delivered to the users, while coded caching demands coordination. The second phase is where they formulate two optimization problems in order to minimize the total energy consumption without affecting the Quality of Service - QoS. On the final phase, numerical results show that uncoded caching can achieve higher energy efficiency rather than coded caching.

## 4. Edge caching in Healthcare

In recent years, more and more scientific papers on healthcare applications presenting the advantages they have with fog technology. Guibert et al. [20], in their study point out that the low delays, that need to be achieved to communicate in emergencies, cannot be achieved through cloud methods. The solution comes through the so-called E-health fogs, where combined with the Content-Centric Network (CCN), integrated health systems can be created and a rapid response to the needs can be achieved. They propose such a health system that they take into account the low delays, the rapid processing of the data and the way the information is stored locally. The basic feature of their own CCN, is that it works with two different types of packets, the first is data and the second is interest for data. Another important feature of their network is that data that is highly demanded by the nodes is stored at strategic points so as to speed up the whole process of communication and response. Ndikumana et al. [21], state that edge architectures consisting of mobile devices face the problem of limiting the computing and storage power of these devices with a consequent delay, especially when the data they manage is medical images. The authors propose an architecture in which edge servers work together to cache the data and then store it. Each server has at its disposal both computer and storage capability to serve the hosts and each host in turn connects to the server with which it has the best communication based on signal strength. The key feature of their proposed architecture is that servers communicate, so that they can make the best use of available resources at any given time. Furthermore, resources available to service requesters, are directly related to the set of requests they have created, as well as to the payments they have made. In another scientific work [22], it is reported that fog technology in relation to cloud technology, is superior to the information transmission speed, also there is not much dependence on the bandwidth of the channel, furthermore better security is achieved, and the computational cost is smaller. It refers extensively to medical monitoring systems for people suffering from diabetes and how these systems benefit from fog architectures. Since data from such devices is small in volume (blood glucose markers), their processing is very easy on fog devices and any decisions are made within a few seconds. There are many cases, where fog devices, are mobile phones with enough processing power and storage capability, but remain battery depended. In the case of the battery depletion, there is no ability of monitoring an event. Such problems, of course, are solved through integrated monitoring systems, for diabetics, connected directly to a fog device.

There are also cases that fog-edge technology is combined with other technologies, in order to achieve the best possible results in medical applications. For example in [23] authors study edge caching in 5G networks and suggest a small cloud architecture and microcell cloud based on SDN based mobile network. They claim that this architecture can be used in mobile health systems for high reliability and mobility. In another work Chen et al. [24] are studying and developing a system for transmitting visual and non-visual medical information through edge caching and Vision Sensing (DVS) technologies. They report the advantages of edge caching, such as the speed of information transmission, especially in hospital environments. They use a camera with dynamic vision sensor which has the ability to record content at a rate in accordance with the moving conditions of the patient. This solution is useful in-patient monitoring. This solution is useful in-patient monitoring. The results of their experiments showed that the use of fog architecture reduces the transmission of information by 36% and using DVS (if visual material is present) up to 86%.

It is evident that the use of fog-edge architectures is increasingly being used in medical data. So, there is a need to study the best way for such a system to develop. Rahmani et al. [25] in their study report how IoT healthcare systems in homes or hospitals can benefit from the use of fog architectures. They propose that a medical monitoring system consists of medical sensors and actuators that receive medical data from patients. The system consists of the gateways that provide the interfaces between the network sensors and the internet, or the local switches and the back-end system, in which medical data is stored for deeper analysis. They analyze the properties of a Fog e-health gateway and report the main pillars of it. One of them, is that there must obviously be local data processing from a fog device near the patient. Medical applications can gain from a preprocessing step that includes data filtering, due to the fact that many times bio-signals are of a particular form and signal processing is necessary prior to their transmission. It is also very important to have adequate compression so that the data can be transmitted faster without introducing alteration. In cases like critical healthcare systems, it is necessary for a health edge system to be able to analyze the data in real time, so that decisions are made immediately, avoiding the need to be transmitted to the cloud The process of data fusion is also an important step, considering that, if the system consists of a large number of sensors, data fusion can decrease the volume of the data, thus decreasing the necessary time for transmission to the network. The above data was used to implement a real-time fog e-health system for recording and transmitting electrocardiograms (EGG) and comparing it with a corresponding cloud-based system. Their experiments showed up to 86.1% latency reduction. In 2017 [26] in their study they propose a network architecture based on small cell 5G networks to transmit medical situations either emergency or non-emergency with the less latency possible. The proposed framework makes use of the known small cells that exist in 3GPP, combining them with edge caching. The above technique constitutes the innovation of the particular proposal. The main purpose of using edge caching, is to transfer the information and measurements of e-health devices to the respective edge nodes and then when traffic and network conditions allow it, to be transmitted to the cloud for basic processing and analysis of the information. In fact, in any given time, an e-health device is chosen to be the basic device to which other devices will be connected to and it will be responsible for transmitting data to the cloud. The choice of the basic device is not accidental but intelligent, as many factors contribute to this decision, such as the percentage of battery, the connection of the device to the network etc. Their suggestion is that small cells can be used with either mmWave, LTE-A or Wi-Fi. Each user who uses an e-health device, such as a wearable device, is the so-called small cell member, where each time it is connected to a corresponding and most suitable cell in order to transmit the information. In addition, downlink edge caching is possible if necessary, something which can be achieved with the increased bandwidth of 5G. In their architecture, e-health devices do not transmit data directly but only when connected to the optimal cell that will also provide the optimal bandwidth. An exception to this, is the case where there is no high utilization of the network, which is the main purpose of this proposal. The same year, authors in [27] developed a fog-based system that enables the storage and communication of various medical data such as phonocardiogram, electrocardiogram and speech data. They support, that with the use of a fog system, they are able to process medical data, within seconds, as opposed to a cloud solution, that may take even several hours until the final stage of the data processing is completed. In their proposed architecture, information flows from the user-patient to the application's processing interfaces. The transmission protocol is the TCP, where TCP sockets are combined with SSL sockets to increase security. The front-end part of the system uses app services through which patients can enter the platform. A backend cloud database is used to permanently store data using MySQL. Through their study and experiments, it is clear that implementing a fog approach, the volume of data that should be transmitted and processed in cloud servers is drastically reduced as most of the information processing is done on fog devices. The authors don’t neglect to acknowledge that similarly to other architectures, fog architecture has also drawbacks. In this case, fog devices have limited processing and storage capabilities, which means that within a few days, the storage space is filled with audio files that occupy many Megabytes up to a few Gigabytes.

As it is obvious, the number of healthcare applications that make use of fog-edge computing or caching is abundant, which has led to the creation of benchmark suites to compare them, such as [28] where they created a benchmark suite for the Internet of Medical Things (IoMT) to help them compare their medical applications and architectures so that every time they will be able to find the optimal microprocessor architecture for each application. This benchmark differs from other open source also benchmarks as it takes account of edge-computing-caching IoMT applications in which edge devices are capable of processing and analyzing information.

## 5. Edge caching in Virtual Reality Videos

Emerging technologies and applications, with higher, and even strict, requirements in data rate, QoS and latency, demand a different approach from what traditional networking has to offer. New smart technologies are becoming more and more computing intensive and require a special methodology, especially when they feature a high level of user interactivity. Such applications are Virtual Reality (VR) and Augmented Reality (AR). While highly interactive VR and AR applications promise to change the way we view the world, such applications require real time-information (e.g. user’s position and direction [29]) that need to be processed and accessed with as little latency as possible [30]. Therefore, it is safe to conclude that these applications mandate new kinds of requirements for their smart technologies that traditional network architectures cannot fulfil.

In [31], authors tackle with problems that arise, when trying to serve virtual reality videos over wireless networks. In a MEC architecture, MEC server delivers the content which is not stored in the VR device, and then the VR device is responsible to construct the task. This results, in keeping the communication cost minimal but increasing the delay. On the other hand, if the MEC server computes the whole task in order to reduce the delay then inevitably increases the communication cost. Authors perform a task scheduling strategy based on Lyapunov theory, to decide which tasks, the MEC server should compute in order to balance the communication cost and the delay, thus taking advantage of the caching resource. In their work, it is proven that the transmission rate is inversely proportional to the computing ability of mobile VR device. In particular they show the correlation between the minimum of average transmission data per task and the computing ability and the caching size of the mobile VR device.

Authors in [32] also in the same context, try to take advantage of the MEC caching and computing capabilities. Their main idea is to cache parts of field of views (FOVS) ahead of time. Afterwards, the necessary processing is run at the VR device. They put together a decision strategy that takes into account both caching and computing parameters, thus enabling them to reduce the transmission rate without sacrificing the response time. In their work they analyze a typical 360 VR video scenario. Among others, they observe that the part of projection has low computational complexity therefore it can be handled by the VR device resulting in reducing the load of the link in half. In general any pre-processing tasks are assigned to be computed in the cloud whereas any post-processing tasks are assigned to be computed at the MEC server. In homogenous FOVs, they prove a mathematical expression that combines communications, caching and computing resources with the ultimate goal to calculate the best policy about caching and computing in the VR device. Similarly they achieve good performance in heterogeneous FOVs.

Content Delivery Networks (CDN) empowered by Edge computing strategies [29], need to adopt new techniques in order to address these new requirements. When it comes to these new requirements, Edge networks provide significant advantages in comparison to traditional networks. These advantages greatly complement the real-time content delivery required by VR and AR applications. The typical architecture where processing and content storage both manifest at the Cloud is not adequate for applications that greatly depend on low-latency. For example, Cloud processing is unacceptable for an AR application where an image must be identified fast. This type of service introduces a latency that is really the sum of two components [30]:

- The Network latency: The data must be transferred to the Cloud in order to be processed but also for the results to be transferred back after processing has finalized.
- Compute Time: The time to process the transferred data.

Content caching at the network edge appears to be the only prevalent solution for low-latency, high-interaction and heavy-throughput applications such as VR and AR [33].

We list some of the advantages that are directly related to the effect of Edge computing and strategic edge caching to interactive computing applications [29]:

- Reduced Latency: Moving the processing and storage caching near the end user has a prolific impact on the performance of highly interactive virtual and augmented reality applications.
- Bandwidth Reduction: By moving the cached content to be processed close to the end user, computing occurs at the edge. Proactive caching and increased cache capacity offer even greater benefits in bandwidth savings.
- High Energy Efficiency: Offloading computing to smaller distributed Data Centers, combined with the equivalent decentralization of storage caching has significant effects in Energy efficiency [34].

To empower Edge computing a new networking paradigm must be adopted. Information Centric Networking (ICN) enhances content delivery by adding new features, such as caching and traffic engineering, which while they are critical for VR and AR applications, they are lacking on today’s networking world [35]. In fact, there is significant evidence in the bibliography to justify that ICN’s provide substantial benefits when it comes to the support of VR and AR applications. When it comes to caching, ICN provides the following two advantages [35]:

- Prefetching of data: On cases where data is more or less static, prefetching and local caching at the edge provides a significant advantage to interactivity [30] complemented by faster processing and bandwidth reductions. As use cases, one may consider VR and AR applications that run on Educational environments, Museums, Real Estate and even Retail.
- Data distribution: Further bandwidth reduction and faster response can be achieved by performing content placement and content distribution at the same time. Furthermore, multicasting can be used to deliver cache content from a single feed [36]. As use cases, one may consider VR and AR applications that run on Sporting events, Computer gaming and public entertainment events such as concerts and theaters.

In regards to mobile networks, caching at the edge of the cellular network provides a number of advantages for the VR and AR applications as well as for the network itself [29]. In addition to the advantages mentioned previously, edge caching prevents the redundant transmissions of content and improve users' overall experience due to the resulting reduced latency. The approach of MEC, is to cache content at the base stations themselves [35] [34], thus providing a highly distributed edge-caching scheme. The critical component in regards to base station caching is the selection of the cache content (i.e. the problem of content popularity) in order to gain the maximum oper-ational efficiency and latency-less content delivery. Considerations for MEC caching are a) Caching place, b) Content popularity, c) Policies and Algorithms [33], d) Content type, e) Mobility awareness, f) Impact on overall system performance [29][38]. Furthermore, aggregate reward earnings can be achieved, by utilizing cooperative caching, rendering and streaming strategies at the base station level, for VR applications [34].

Regardless of the environment and methodology of delivery, VR and AR applications essentially require a new network architecture that will be able to accommodate fast dynamic multicast delivery of cache content closer to the user where it is needed [36]. In regards to the content, strategic selection [33] must be done in order to cache the content that is needed the most. Content that is requested by the user will be cache via the reactive caching strategy while at the same time the system will be able to utilize prediction techniques in order to fetch and cache content before it is requested by the user. This later caching strategy is called proactive strategy and it can be achieved by utilizing a number of variables such as mobility, locality but also social traffic analysis, thus anticipating the user's upcoming requests [36]. Furthermore, proactively cached content can be processed utilizing the predictions based on application use and therefore the computed content can now be cached. This is a great benefit to VR and AR applications as they rely on fast content provision, crucial for fast interaction, and effective immersive application delivery [36], [37].

## 6. Conclusion

The growth of devices that connect to the internet, has established a need for uninterrupted, lossless and with minimal delay, communications. The high volume data traffic in combination with low latency has become a mandatory requirement for many emerging trends.

This paper presented a brief literature review of applications of Edge Caching, in order to prove the importance of Edge caching methods in areas such as Smart Cities, Vehicular Content Networks, Healthcare and Virtual Reality Videos. In all cases, the main goal is to achieve minimal communication cost under the low delay constrain. Content caching at the network edge appears to be the only prevalent solution for low-latency, high-interaction and heavy-throughput applications because edge caching achieves reduced latency, bandwidth reduction and high energy efficiency.

## 7. References

[1] H. Zhu, Y. Cao, X. Wei, W. Wang, T. Jiang and S. Jin, "Caching Transient Data for Internet of Things: A Deep Reinforcement Learning Approach," *IEEE Internet of Things Journal,* 2018.

[2] A. Orsino, R. Kovalchukov, A. Samuylov, D. Moltchanov, S. Andreev, Y. Koucheryavy and M. Valkama, "Caching-Aided Collaborative D2D Operation for Predictive Data Dissemination in Industrial IoT," *IEEE Wireless Communications,* vol. 25, no. 3, pp. 50-57, 2018.

[3] F. Al-Turjman, "Cognitive caching for the future sensors in fog networking," *Pervasive and Mobile Computing,* vol. 42, pp. 317-334, 2017.

[4] F. Song, Z.-Y. Ai, J.-J. Li, G. Pau, M. Collotta, I. You and H.-K. Zhang, "Smart collaborative caching for information-centric IoT in fog computing," *Sensors,* vol. 17, no. 11, p. 2512, 2017.

[5] M. Meddeb, A. Dhraief, A. Belghith, T. Monteil, K. Drira and S. AlAhmadi, "Cache freshness in named data networking for the internet of things," *The Computer Journal,* vol. 61, no. 10, pp. 1496-1511, 2018.

[6] Y. Wei, F. R. Yu, M. Song and Z. Han, "Joint Optimization of Caching, Computing, and Radio Resources for Fog-Enabled IoT Using Natural Actor-Critic Deep Reinforcement Learning," *IEEE Internet of Things Journal,* 2018.

[7] Y. Hao, Y. Miao, Y. Tian, L. Hu, M. S. Hossain, G. Muhammad and S. U. Amin, "Smart-Edge-CoCaCo: AI-Enabled Smart Edge with Joint Computation, Caching, and Communication in Heterogeneous IoT," *arXiv preprint arXiv:1901.02126,* 2019.

[8] Y. He, F. R. Yu, N. Zhao, V. C. M. Leung and H. Yin, "Software-defined networks with mobile edge computing and caching for smart cities: A big data deep reinforcement learning approach," *IEEE Communications Magazine,* vol. 55, no. 12, pp. 31-37, 2017.

[9] Z. Su, Y. Hui, Q. Xu, T. Yang, J. Liu and Y. Jia, "An edge caching scheme to distribute content in vehicular networks," *IEEE Transactions on Vehicular Technology,* vol. 67, no. 6, pp. 5346-5356, 2018.

[10] S. Zhang, W. Quan, J. Li, W. Shi, P. Yang and X. Shen, "Air-ground integrated vehicular network slicing with content pushing and caching," *IEEE Journal on Selected Areas in Communications,* vol. 36, no. 9, pp. 2114-2127, 2018.

[11] A. Mahmood, C. E. Casetti, C. F. Chiasserini, P. Giaccone and J. Härri, "The RICH Prefetching in Edge Caches for In-Order Delivery to Connected Cars," *IEEE Transactions on Vehicular Technology,* vol. 68, no. 1, pp. 4-18, 2019.

[12] Y. He, N. Zhao and H. Yin, "Integrated networking, caching, and computing for connected vehicles: A deep reinforcement learning approach," *IEEE Transactions on Vehicular Technology,* vol. 67, no. 1, pp. 44-55, 2018.

[13] B. Hu, L. Fang, X. Cheng and L. Yang, "In-Vehicle Caching (IV-Cache) Via Dynamic Distributed Storage Relay (D ^ 2 SR) in Vehicular Networks," *IEEE Transactions on Vehicular Technology,* vol. 68, no. 1, pp. 843-855, 2019.

[14] H. Moustafa, E. M. Schooler and J. McCarthy, "Reverse cdn in fog computing: The lifecycle of video data in

connected and autonomous vehicles," in *2017 IEEE Fog World Congress (FWC)*, 2017.

[15] M. Li, P. Si and Y. Zhang, "Delay-tolerant data traffic to software-defined vehicular networks with mobile edge computing in smart city," *IEEE Transactions on Vehicular Technology,* vol. 67, no. 10, pp. 9073-9086, 2018.

[16] G. Tanganelli, C. Vallati and E. Mingozzi, "A fog-based distributed look-up service for intelligent transportation systems," in *2017 IEEE 18th International Symposium on a World of Wireless, Mobile and Multimedia Networks (WoWMoM)*, 2017.

[17] Z. Yan, M. Peng and C. Wang, "Economical energy efficiency: An advanced performance metric for 5G systems," *IEEE Wireless Communications,* vol. 24, no. 1, pp. 32-37, 2017.

[18] T. X. Vu, S. Chatzinotas, B. Ottersten and T. Q. Duong, "Energy minimization for cache-assisted content delivery networks with wireless backhaul," *IEEE Wireless Communications Letters,* vol. 7, no. 3, pp. 332-335, 2018.

[19] P. Liu, G. Xu, K. Yang, K. Wang and X. Meng, "Jointly Optimized Energy-Minimal Resource Allocation in Cache-Enhanced Mobile Edge Computing Systems," *IEEE Access,* vol. 7, pp. 3336-3347, 2019.

[20] D. Guibert, J. Wu, S. He, M. Wang and J. Li, "CC-fog: Toward content-centric fog networks for E-health," in *2017 IEEE 19th International Conference on e-Health Networking, Applications and Services (Healthcom)*, 2017.

[21] A. Ndikumana, S. Ullah, T. LeAnh, N. H. Tran and C. S. Hong, "Collaborative cache allocation and computation offloading in mobile edge computing," in *2017 19th Asia-Pacific Network Operations and Management Symposium (APNOMS)*, 2017.

[22] D. C. Klonoff, *Fog computing and edge computing architectures for processing data from diabetes devices connected to the medical Internet of things,* SAGE Publications Sage CA: Los Angeles, CA, 2017.

[23] M. Chen, Y. Qian, Y. Hao, Y. Li and J. Song, "Data-driven computing and caching in 5G networks: Architecture and delay analysis," *IEEE Wireless Communications,* vol. 25, no. 1, pp. 70-75, 2018.

[24] Z. Chen, T. Shikh-Bahaei, P. Luff and M. Shikh-Bahaei, "Edge caching and Dynamic Vision Sensing for low delay access to visual medical information," in *2017 39th Annual International Conference of the IEEE Engineering in Medicine and Biology Society (EMBC)*, 2017.

[25] A. M. Rahmani, T. N. Gia, B. Negash, A. Anzanpour, I. Azimi, M. Jiang and P. Liljeberg, "Exploiting smart e-Health gateways at the edge of healthcare Internet-of-Things: A fog computing approach," *Future Generation Computer Systems,* vol. 78, pp. 641-658, 2018.

[26] A. Radwan, M. F. Domingues and J. Rodriguez, "Mobile caching-enabled small-cells for delay-tolerant e-Health apps," in *2017 IEEE International Conference on Communications Workshops (ICC Workshops)*, 2017.

[27] H. Dubey, A. Monteiro, N. Constant, M. Abtahi, D. Borthakur, L. Mahler, Y. Sun, Q. Yang, U. Akbar and K. Mankodiya, "Fog computing in medical internet-of-things: Architecture, implementation, and applications," in *Handbook of Large-Scale Distributed Computing in Smart Healthcare*, Springer, 2017, pp. 281-321.

[28] A. Limaye and T. Adegbija, "HERMIT: A benchmark suite for the internet of medical things," *IEEE Internet of Things Journal,* vol. 5, no. 5, pp. 4212-4222, 2018.

[29] S. Wang, X. Zhang, Y. Zhang, L. Wang, J. Yang and W. Wang, "A Survey on Mobile Edge Networks: Convergence of Computing, Caching and Communications," *IEEE Access,* 2017.

[30] U. Drolia, K. Guo, J. Tan, R. Gandhi and P. Narasimhan, "Cachier: Edge-Caching for Recognition Applications," in *Proceedings - International Conference on Distributed Computing Systems*, 2017.

[31] X. Yang, Z. Chen, K. Li, Y. Sun, N. Liu, W. Xie and Y. Zhao, "Communication-constrained mobile edge computing systems for wireless virtual reality: Scheduling and tradeoff," *IEEE Access,* vol. 6, pp. 16665-16677, 2018.

[32] Y. Sun, Z. Chen, M. Tao and H. Liu, "Communications, Caching and Computing for Mobile Virtual Reality: Modeling and Tradeoff," *arXiv preprint arXiv:1806.08928,* 2018.

[33] Y. Liu, Z. Han, H. Cao, F. Li, G. Li, Q. Shen and J. Li, "Data-driven Approaches to Edge Caching," 2018.

[34] J. Chakareski, "VR/AR Immersive Communication: Caching, Edge Computing, and Transmission Trade-Offs," in *Proceedings of the Workshop on Virtual Reality and Augmented Reality Network - VR/AR Network '17*, 2017.

[35] C. Westphal, "Challenges in Networking to Support Augmented Reality and Virtual Reality," *ICNC 2017,* 2017.

[36] E. Bastug, M. Bennis, M. Medard and M. Debbah, "Toward Interconnected Virtual Reality: Opportunities, Challenges, and Enablers," IEEE Communications Magazine, 2017.

[37] Nikoloudakis, Y., Markakis, E., Alexiou, G., Bourazani, S., Mastorakis, G., Pallis, E., ... & Mavromoustakis, C. (2018, September). Edge Caching Architecture for Media Delivery over P2P Networks. In 2018 IEEE 23rd International Workshop on Computer Aided Modeling and Design of Communication Links and Networks (CAMAD) (pp. 1-5). IEEE.

[38] Markakis, E. K., Karras, K., Sideris, A., Alexiou, G., & Pallis, E. (2017). Computing, caching, and

communication at the edge: The cornerstone for building a versatile 5G ecosystem. IEEE Communications Magazine, 55(11), 152-157.

[39] Markakis, E., Negru, D., Bruneau-Queyreix, J., Pallis, E., Mastorakis, G., & Mavromoustakis, C. X. (2016). A p2p home-box overlay for efficient content distribution. In Emerging Innovations in Wireless Networks and Broadband Technologies (pp. 199-220). IGI Global.